# Lupascian Non-Negativity Applied to Conceptual Modeling: Alternating Static Potentiality and Dynamic Actuality

Sabah Al-Fedaghi*
*Computer Engineering Department*
*Kuwait University*
*Kuwait*
salfedaghi@yahoo.com, sabah.alfedaghi@ku.edu.kw

*Abstract* - **In software engineering, conceptual modeling focuses on creating representations of the world that are as faithful and rich as possible, with the aim of guiding the development of software systems. In contrast, in the computing realm, the notion of ontology has been characterized as being closely related to conceptual modeling and is often viewed as a specification of a conceptualization. Accordingly, conceptual modeling and ontology engineering now address the same problem of representing the world in a suitable fashion. A high-level ontology provides a means to describe concepts and their interactions with each other and to capture structural and behavioral features in the intended domain. This paper aims to analyze ontological concepts and semantics of modeling notations to provide a common understanding among software engineers. An important issue in this context concerns the question of whether the modeled world might be stratified into ontological levels. We introduce an abstract system of two-level domain ontology to be used as a foundation for conceptual models. We study the two levels of staticity and dynamics in the context of the thinging machine (TM) model using the notions of potentiality and actuality that the Franco-Romanian philosopher Stéphane Lupasco developed in logic. He provided a quasi-universal rejection of contradiction where every event was always associated with a 'no event,' such that the actualization of an event entails the potentialization of a 'no event' and vice versa without either ever disappearing completely. This approach is illustrated by re-modeling UML state machines in TM modeling. The results strengthen the semantics of a static versus dynamic levels in conceptual modeling and sharpen the notion of events as a phenomenon without negativity alternating between the two levels of dynamics and staticity.**

## I. Introduction

In software engineering, conceptual modeling focuses on creating representations of the world that are as faithful and rich as possible, with the aim of guiding the development of better software systems. According to Gonzalez-Perez [1], "a conceptual model, in a nutshell, is a representation of a portion of the world that is made of concepts. All this happened in the context of software engineering." In contrast, in the computing realm, the notion of *ontology* has been characterized as being closely related to conceptual modeling and is often used as a tool for the specification of a conceptualization [2]. Accordingly, ontologies and conceptual models are not that different. According to [1], "some authors claim now that ontologies are just a particular kind of conceptual models, and a new field of ontology engineering has arisen to take much of the work in ontologies and combine it with some advances in conceptual modeling"

Accordingly, conceptual modeling and ontology engineering now address the same problem of representing the world in a suitable fashion. An ontology addresses real-world descriptors of entities and is thus better named "domain ontology" [3]. Domain ontology provides means to describe concepts and their interactions with each other in the intended domain and to capture the modeled system's structural and behavioral features. In this paper, we explore an ontology that supports concepts and semantics of modeling notations in order to provide a common understanding of those concepts among software engineers.

An important issue in this context concerns the question of whether the world might be stratified into ontological levels. For example, the Eastern conceptualization of Yin and Yang is a type of levelization of interest in our work. Yin is the source related to negativity and passiveness, which would be the static state of things. Yang is related to behavior, which is regarded as positive and active; this would be the dynamic state of things. The issue here is that things and actions *per se* ontologically *alternate* between potentiality and actuality. Levels have been mentioned in both philosophical and scientific publications [4].

According to List [5], we explain the phenomena in a modeled domain of interest by focusing on *higher-level* properties and regularities. Some philosophers ask whether the world might be *layered* or *stratified into levels*. The levels in question are not just levels of description but ontological levels [5]. Among these philosophers, Stéphane Lupasco (1900–1988) described a logical system applicable to real processes at higher levels of reality [6]. Lupasco's ideas involve that, in the presence of any phenomenon, what its contradictory phenomenon is and second, "to what extent it potentializes it or is potentialized by it" [6]. According to Judge [7], for Lupasco, all human cognitive and practical efforts oscillate between extension and intensity. For the human being, extension is that which one knows more than one feels, whereas intensity is that which one feels more than one knows. The characteristics of each of these notions recall recent work on right and left hemispheres of the brain [7].

---
*Retired June 2021, seconded fall semester 2021/2022

In our case, we will utilize a conceptual model with two-level domain ontology based on a conceptualized static world (potentiality) with its implemented dynamic reality (actuality). The notions of potentiality and actuality sharpen the distinction between the two levels of modeling. The results strengthen the semantics of concepts such as static descriptions and actions, static entities, and dynamic specifications involving events and behavior. Additionally, Lupasco's ideas can assist in modeling negativity actions.

*A. This Paper*

In this paper, we apply an abstract system of two-level domain ontology to be used as a foundation for conceptual models in software engineering. We study these two levels of abstraction in the context of the thinging machine (TM) model (see Section 2) applied to UML examples. We emphasize contrasting TM static actions with behavior based on events.

An abstraction is a crucial element in computer science, and it takes many different forms. Some argue that abstraction in computer science fundamentally differs from abstraction in mathematics because it "must leave behind an implementation trace" [8]. Computational abstraction depends on the existence of an implementation, for example, "even though [UML] classes hide details of their methods [operations], they must have implementations. Hence, computational abstractions preserve both an abstract guise and an implementation" [8].

*B. Related Problems*

The research problem in this paper aims to study how to relate the structural and dynamic levels in the specification of a conceptual model. These two levels also involve static entities and dynamic actions. We aim to develop an ontological mapping from staticity to dynamics, for example, as in relating UML class and state diagrams. Typically such an issue focuses on notations, such as UML sequence diagrams, class and activity diagrams, data-flow diagrams, Petri-nets, state machines and charts, and message sequence charts, that accommodate some or all of the behavior expressed in functional requirements and designs [9]. UML involve the data's structural level that the system processes and behavioral level where we model the system's dynamic behavior and how it responds to events.

The UML representations of structure and behavior overlap, raising consistency and integration problems. Moreover, the object-oriented nature of UML set the foundation for several behavioral views in UML, each of which is a different alternative for representing behavior [10][11]. Alternatively, the materials of this paper strengthen the foundation of the TM modeling and shows that the TM description exhibits clarity of semantics with level distinction between the static and dynamic models. For contrasting purposes, at the dynamic level, we emphasize the UML state machine representation of behavior. State analysis is typically used for modeling a system's behavior and to specify the control design and operations for this system [11].

*C. Sections of the Paper*

Section 2 presents a brief review of the TM modeling process that was previously developed (See a recent paper [12]). Section 3 includes a sample TM modeling of a washing machine described using a UML state machine diagram. Section 3 includes an analysis of modeling that described potential (static) actions and actualities represented by events. We take these ideas from the philosopher Stéphane Lupasco, who provided a theoretical basis for the quasi-universal rejection of contradiction where very event is always associated with an anti-event, such that the actualization of an event entails the potentialization of a non-event and *vice versa*. Section 4 focuses on the notion of behavior as actuality. Section 5 presents a case study of applying potentiality and actuality to behavior.

## II. THING MACHINE (TM) MODEL

TM modelling is based on one category of entities called thimacs (*thi*ngs/*mac*hines). The thimac has a dual mode of being: the machine side and the thing side. The machine, called TM machine, has the (potential) actions shown in Fig. 1. The term 'machine' reflects the thimac's universal implication that every 'thing' is a machine. The sense of 'machinery' originated in the TM actions indicating that everything that creates, changes (processes) and moves (release-transfer-receive) other things is a machine. Simultaneously, what a machine creates, processes (changes), releases, transfers and/or receives is a thing.

Assemblages of thimacs may be formed from a juxtaposition of subthimacs that are bonded into a structure at a higher level at which they become parts. Thimacs comprise parts, which themselves are thimacs that comprise parts, and so on. Thimacs cannot be reduced to their parts because they (as wholes) have their own machines.

In general, we can view a thimac as a replacement to the notion of *system*, and the machine in Fig. 1 is a generalization of the famous input-process-output machine. We can substitute 'thimac' for 'system' to achieve a further general view of systems. For example, in Kim's [13] description of systems, "You are a member of many systems.

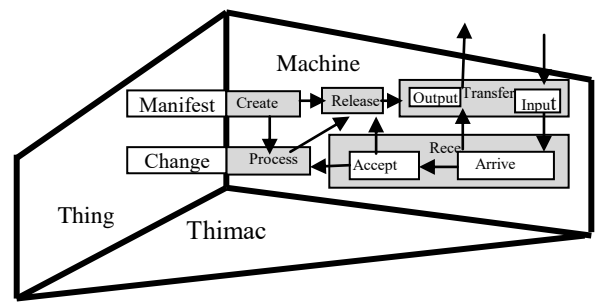

Fig. 1. The thimac as a thing and machine.

You yourself are a complex biological system comprising many smaller systems. And every day, you probably interact with dozens of systems, such as automobiles, ATM machines, etc." We can replace every 'system' in the quote with 'thimac.' For Kim [13], a system is any group of interacting, interrelated or interdependent parts that form a complex and unified whole that has a specific purpose. A thimac is a whole that includes subthimacs that interrelate with each other and interact with the outside through the flow of things (thimacs).

The TM modeling involves two levels that are characterized by staticity and dynamics. Staticity refers to a static model that represents the world of potentialities outside time where everything is present simultaneously. The dynamic model represents the world of actuality in time where events do not necessarily happen simultaneously (see Fig. 2). Time is a thimac (i.e., a thing that creates, processes, releases, transfers and/or receives other things). The two TM models share regions (subthimacs). In a very rough analogy, TM staticity and dynamics parallel universality and particularity, respectively. However, TM staticity and dynamics are applied not only to things but also to actions, thus we have static (potential) *action* in the static model and events (actual *actions*) in the dynamic TM model.

*A. The Machine*

TM actions seen in Fig. 1 can be described as follows.

**Arrive**: A thing moves to a machine.
**Accept**: A thing enters the machine. For simplification, we assume that arriving things are accepted; therefore, we can combine **arrive** and **accept** stages into the **receive** stage.
**Release**: A thing is ready for transfer outside the machine.
**Process**: A thing is changed, handled and examined, but no new thing results.
**Create**: A new thing "comes into being" (is found/manifested) in the machine and is realized from the moment it arises (emergence) in a thimac. Note that for simplicity's sake, we omit *create* in some diagrams, assuming the box representing the thimac implies its existence (in the TM model).
**Transfer**: A thing is input into or output from a machine.

Additionally, the TM model includes a *triggering* mechanism (denoted by a dashed arrow in this article's figures), which initiates a (non-sequential) flow from one machine to another. Multiple machines can interact with each other through the movement of things or through triggering. Triggering is a transformation from the movement of one thing to the movement of a different thing. The TM 'space' is a structure of thimacs that forms regions of events (to be defined later).

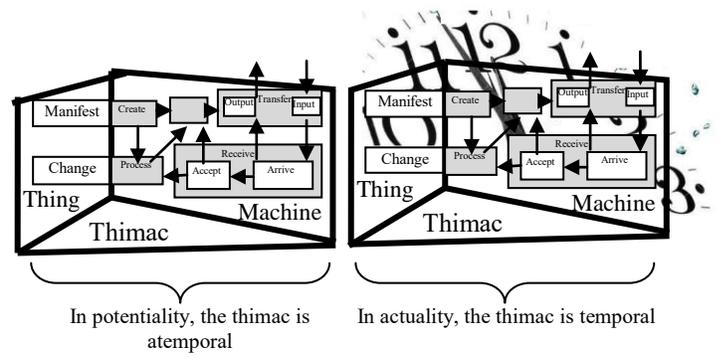

In potentiality, the thimac is atemporal    In actuality, the thimac is temporal

Fig. 2. Illustration of the thimac in potentiality and actuality worlds.

*B. TM Modeling Example*

Consider the state machine given in Sparx Systems [14] shown in Fig. 3. According the Sparx tutorial [14], a UML state machine diagram models through specifying two notions:
(i) the *behavior* of a single object, and
(ii) the *sequence of events* that an object goes through during its lifetime in response to events.

In Fig. 3, a filled black circle denotes the initial state. A circle with a dot inside it denotes the final state. Lines with arrowheads denote *transitions* from one state to the next. In a state machine, an *event* is one of the causes of the transition. Other possible causes are a signal, a change in some condition or the passage of time. Other notions that are used in this context are *guard*, **signal** and *effect*. This state machine model will be contrasted with its corresponding TM model.

Fig. 4 shows the static TM model. The description represents the static domain of potential static things and actions that can become actual manifestations in a particular time, which will be discussed in the dynamic TM model. The static thimac is similar to a form; in metaphysical terms, it is the description of a thimac: its machines and its subthimacs.

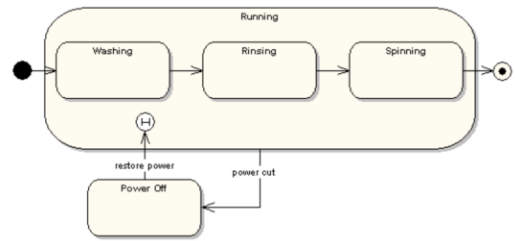

Fig. 3. State machine sample (from [14])

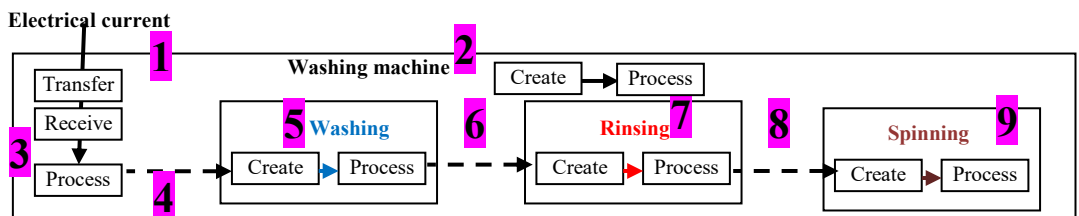

Fig. 4 The static model

In Fig. 4, first, the electrical current (pink number 1) flows into the washing machine (2) to be processed (3—e.g., consumed as energy). This processing of current triggers (4) the washing process (5), which in turn triggers (6) the resining process (7) that triggers (8) the spinning process (9)

This static model is an abstraction that reflects a potentiality (pure capacity) domain of what will become manifested entities and active actions in the next level of domain of actuality (dynamic level) because of bonding with time. Actuality involves temporal stability as part of the world's structure and time. The passage from actuality to potentiality and vice versa is a fundamental concept in TM modeling that we will emphasize in this paper. Dragons and unicorns are static thimacs; horses and lizards are also static thimacs, but unlike dragons and unicorns, they can become instances in the dynamic domain.

The dynamic level includes the same diagram as in Fig. 4 superimposed with regions (subdiagrams of the static model) of events and instances (objects). Events and instances refer to *particular* events and things.

A TM thing is what is created, processed, released, transferred and/or received. A thing has always a *create action* in the static description. Space, in the classical sense, is also a TM thimac. If it is important to express space in the static model, for example, an object is in space, a particular spot of space and so on, then space can be represented as a thimac; that is, space can be created, processed and so forth.

Fig. 5 shows two events.
(i) The event: *The washing machine receives electrical current* (in the left side of the figure). This event is a specific occurrence in time over the indicated region. The region is a subdiagram of the static diagram, that is, Fig. 4.
(ii) The instance (object) of *a particular washing machine* (the right side of the figure). This instance refers to the *existence* of a *particular* washing machine extended in time (a thimac stretches in time). The instance can be viewed as a special type of an event, an extended event. Note that in TM, the existence notion refers to existence in the TM model, that is, not necessarily in space.

Accordingly, we identify the following events as specified in the dynamic model of Fig. 6 where, for simplicity's sake, the event's region represents it.

$E_1$: There exists a washing machine (this machine). Events require the washing machine. What can be washing, rinsing and spinning if there is no particular machine?
$E_2$: Electrical current applied to the machine.
$E_3$: A washing process starts.
$E_4$: The washing process takes its course.
$E_5$: A rinsing process starts.
$E_6$: The rinsing process takes its course.
$E_7$: A spinning process starts.
$E_4$: The spinning process takes its course.

Note it is possible to specify the events in terms of the generic actions (i.e., create, process, etc.). However, this creates many events, each of which corresponds to genetic action, and it is better to specify larger 'meaningful' events when appropriate.

Fig. 7 specifies the behavior model in terms of the chronology of events.

### III. TWO-LEVEL OF REPRESENTATION

The TM modeling is based on distinguishing two worlds: potentialities (also called staticity to avoid any human-related notions) described in terms of static actions (do not embed time) and actualities (dynamics) represented by (temporal) events. This implies that the activation of an event (actions, e.g., process) and its region (e.g., stop process) can be defined as an alteration between the static level and the dynamic level. Instead of 'process' versus 'do not process,' we have 'process' (which moves the action to the dynamic level) versus 'revert to static process,' which returns the process to the static level. We take this method of eliminating negativity from the philosopher Stéphane Lupasco, as we will describe next.

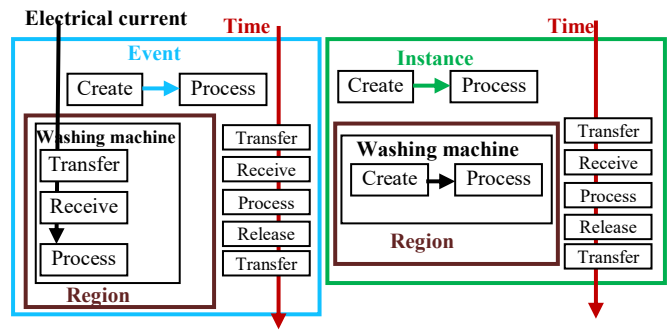

Fig. 5 Examples of an event (left) and instance (right)

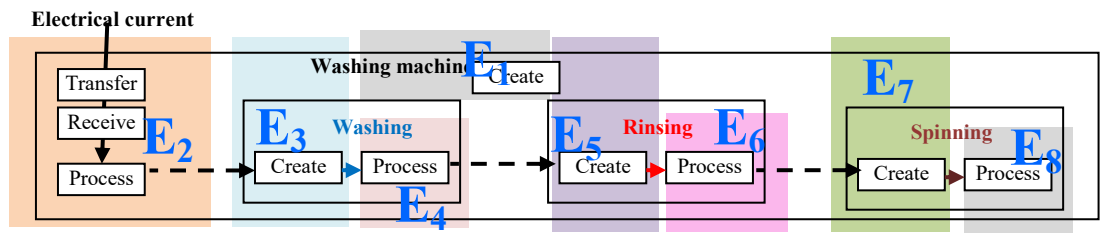

Fig. 6 The dynamic

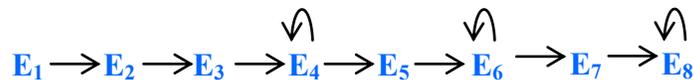

Fig. 7 The behavior model

According to Brenner [15], every element *e* (in TM: an event, i.e., a thimac that contains a region plus time) always associated with a *non-e* (in TM: static thimac), such that the actualization of one entails the potentialization of the other and vice versa, alternatively, "without either ever disappearing completely" [15]. The philosopher Stéphane Lupasco provided a theoretical basis for the quasi-universal rejection of contradiction where very event is always associated with an anti-event, such that the actualization of an event entails the potentialization of a non-event and *vice versa*. These concepts are illustrated in Fig. 8. This theoretical base is provided to emerge from a level of contradiction to the level of time attunement. In Aristotelian language, the alternation involves passage from potency (akin to potentiality) to act.

To illustrate the idea of event/non-event, consider the following example. According to Sloan [16], a light switch that gets flipped has two different events (gets flipped up, gets flipped down) – see Fig. 9. In a UML state diagram, each possible event that can happen to cause an object to change from one state to another is represented by an arrow from the original state to the resulting state, labeled with the name of the event. We will try to understand the notion of *state* applied in this example in terms of "light on" and "light off". Lupasco's concept rejects the existence of the "light off" state.

Fig. 10 shows the static TM model that corresponds to a bulb that receives electrical current and causes the generation of light. It involves three TM events (one of them is of the *instance type*), as shown in Fig 11. In Fig. 11, there are three events, $E_1$: *A bulb exists*, $E_2$: *Current flows* and $E_3$: *Light is created*.

In the figure, the current arrival to the bulb creates the light and such a process continues. The *state* of 'light on' resulted from something *happens*, that is, the (dynamic) flow of electrical current. The question is: What is the '*happening*' that produces the *state* of the 'light off?' Does *nothing* happens (no current) cause 'light off?' This issue is related to the study of events and the relationship between events and states. If the state machine can change from one state to another in response to some inputs, then what causes a bulb to be in an off state when there is no current? In fact, it is in the 'off state' by itself even when there is no switch or if the switch does not work. Therefore, is 'no input (current)' an input? Of course, we can claim that pressing a switch causes the current flow to stop, which in turn causes the light to disappear. Still, *'no current' occurrence* causes the light off. The semantics of 'bulb off' is the negative event 'no current.'

As shown in Fig. 12, 'light off' precedes and succeeds the 'light on' event. If the state is all possible system *occurrences*, then in this example, there is only one occurrence: the electrical current flows. "No current" is not an occurrence, which is the negative of occurrence. The negative of an occurrence is 'no event' in TM. It seems paradoxical that no event is an event. In this case, we use Lupasco's 'no event' to claim that 'no event' reverts from the dynamic level to the static level, an occurrence (at the dynamic level) that we will denote with '$NE_i$', that is, the event $E_i$ returns to potentiality to become only a region.

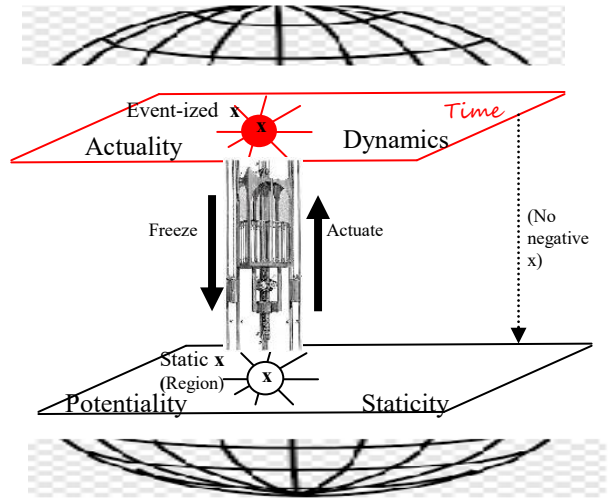

(a) Event and anti-event (Lupasco's language)

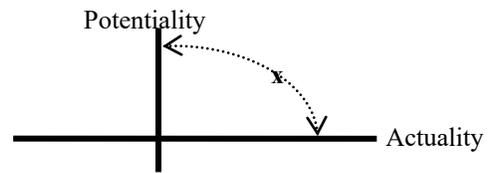

(b) Alteration between potentiality and actuality

Fig. 8 Illustration of event and region, and alteration between potentiality and actuality

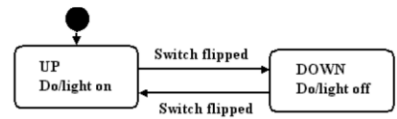

Fig. 9 Two-state machine (from [16])

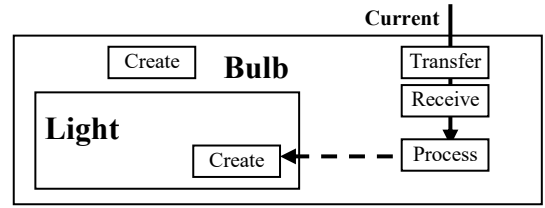

Fig. 10 Electrical current generates light

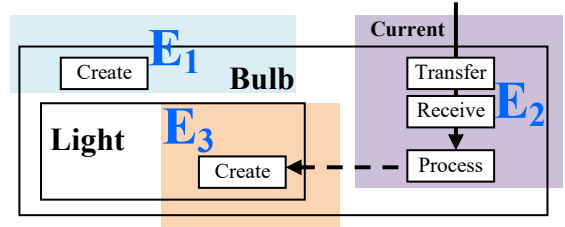

Fig. 11 The dynamic model of electrical current that generates light

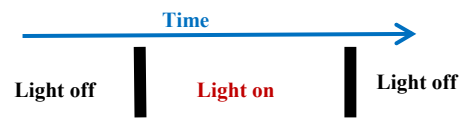

Fig. 12 The event 'light on' preceded and succeeded by 'light off'

P a g e | 5

In state machines, sometime the notion of state is described as the condition of the system at a given time (e.g., in chemistry). 'Condition' mostly means a state of being. In the bulb example, the being can be interpreted as 'light on' and 'light off'. Therefore, here, 'light off' may mean 'initially' and 'after light on,' which can be interpreted as before or after the event of 'light on.'

The bulb example raises the issue of the difference between event and 'no event,' for example, the electrical current 'stops' after it flows. Note that $E_1$ in Fig. 11 is an instance (*there exists a bulb*). This means that its duration continues along other events as illustrated in Fig. 13. In Fig. 13, P denotes potentiality (symbolized as -) and A denotes actuality (symbolized as +). 'Light off' in TM means reverting to potentiality (static case). That is, the states of *on* and *off* reflect only, in Lupasco's words [16], "an alternating of actualizations and potentializations." Thus, the table of Fig. 13, under the assumption of one time slot for all events, shows $E_1$ (the bulb) is created and persists in an *extended existence*. Extendibility refers to how an object can be wholly present now, yet also wholly present at a later time. We *postulate* that such extendibility *may* refer to stretching out the thimac as exemplified in Fig. 14. Such a topic is not of concern here and we will leave it to future research.

In Fig. 13, $E_2$ and $E_3$ become events then subsided. When there is no flow of current, then reverting to potentiality (disappearance of the time ingredient) is called "light off." Accordingly, Fig. 15 shows the behavior model of the bulb. Fig. 16 shows the resultant situation where $E_1$ and $E_2$ disappear from the actuality theater. Disposing of negativity is advantageous in some modeling situations, which we will demonstrate later in this paper.

The conclusion to which we are driving is that *the stoppage of an event at a certain region is <u>a return to staticity</u> in that region.* Accordingly, in TM, there are only two possibilities: a static system and dynamic system (roughly, sometimes called 'intensional' and 'extensional,' respectively). A thing (thimac) disappears from the dynamic picture, meaning that is loses its time component and becomes a region again instead of being part of an event.

IV. POTENTIALITY AND ACTUALITY IN TM

The difference between the two TM modelling levels—potential (static) and actual (dynamic)—is time. At the potentiality level, the TM static model is the description of the domain's structure. It includes everything (thimacs) that creates/'being created,' processes/'being processed,' releases/'being released,' transfers/'being transferred' and receives/'being received'. For example, a human being as a thimac creates, processes (changes), releases, transfers and receives things. Simultaneously, the human being is created (e.g., being born), being processed (e.g., change hair), being released (e.g., from prison), being transferred (e.g., between places) and being received (e.g., in a university).

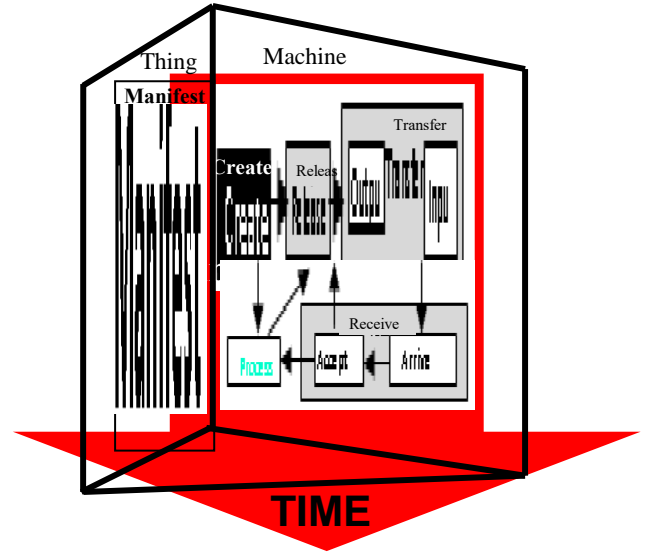

Fig. 13 Durations of different TM events

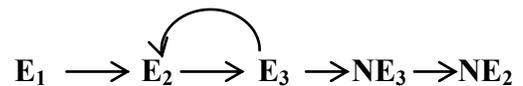

Fig. 14 Illustration of an instance that is extended in time.

$$E_1 \longrightarrow E_2 \longrightarrow E_3 \longrightarrow NE_3 \longrightarrow NE_2$$

Fig. 15 The behaviour model of electrical current that generates light

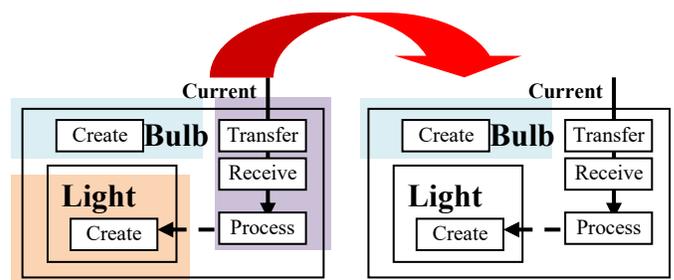

Fig. 16 Stopping events means returning to the static model

A 'static' action refers to potential action, that is, the principles of momentum. For example, take the static action *calculate* as a part of a computer program listing; the action *calculate* becomes an actual calculation action when the program is in the execution phase.

At the actuality level, instances and events unfold dynamically in time, constructed from the unity of regions in the static model and time. Regions are subdiagrams of the static model. An *instan*ce of a thing (thimac) exists only in actuality while its static region is always unobservable in actuality and present in potentiality. Similarly, an action exists (activate, e.g., executed calculate in a computer program) only in actuality while its static version *presents* in potentiality.

Potentiality (e.g., create walking) resides in the TM static model. Actuality arises in the dynamic TM model, assuming that the corresponding static model includes its potentiality. The dynamism is the actuality of a thing that has potentially. This discussion parallels Aristotle's ideas in this context [17].

We just recast the idea of alternating actualizations and potentializations (see Brenner's [15] ideas about potentialization and actualization, the so-called logic in reality [LIR]) in the context of conceptual modeling and systems behavior. A TM event implies that the thimac's region is in actuality and that 'stoppage of the event' means to revert the region to its original potentiality (static) level.

Take an example from Brenner [15] that concerns the presence and the absence of a black dot, or a change of state, from 'being no black dot at all' to 'being one,' modeled in Fig. 17. In Fig. 17, first, the dot is a potentiality (static thimac), then it is an instance event, and then it reverts to potentiality when being erased. Accordingly, the occurrence and stoppage of an event can be conceptualized as alternating between actualizations and potentializations. This means that there is no negative implication in the classical sense; hence, as Brenner expressed [15], "the actualization of one [e.g., logical expression p] entails the potentialization of the other [not p]."

## V. Behavior as Actuality

The two-level modeling separates the systems dynamics from staticity. In general, dynamic means capable of action and/or change, while static means stationary or fixed. The static domain is a frozen, unchanging whole containing all things [18]. In this sense, the static domain does not grow or shrink; it does not lose what is past and gain what is future [19] (note that, in this sentence, we have replaced "reality" in [19] with 'static domain'). Next, we focus on the dynamic level, specifically on the notion of behavior, for which academics rarely provide an explicit definition of what 'behavior' actually is [20].

Behavior is a key entity in understanding many systems' driving forces. It has also been increasingly highlighted for complex problem solving within virtual and physical organizations, in particular for pattern analysis, business intelligence, social computing, web usage and network monitoring [21]. Behavior is a core concept in UML and UML behavioral diagrams (use case, activity, interaction, state machine, protocol state machine, OCL) define how the UML represents "resources interaction and how they execute their capabilities" [22]. Specifically, in this paper, we focus on behavior that is modeled using *state machines* as in the following case study.

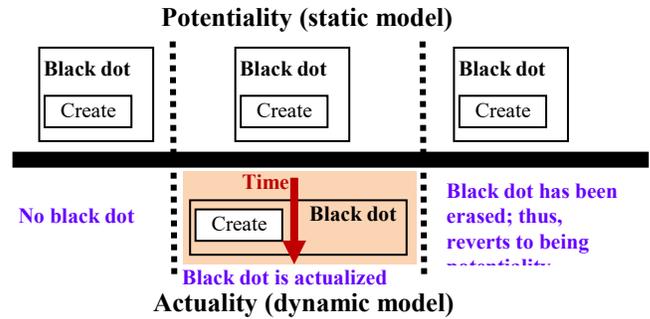

Fig. 17 Illustration of a black dot that 'exists' and is then erased

According to Winter et al. [23], in engineering, a set of requirements is understood as what is needed as a system's behavior from the customer's perspective. In their case study, Winter et al. [23] assume requirements are given as natural language descriptions and focus on modelling the requirements (i.e., building a specification), analysis and verifying the resulting model.

Winter et al. [23] provide an example of requirements for the *Microwave Oven* as published in [9] in Fig. 18 (left). Winter et al. [23] use behavior trees in specifying the oven's behavior. The behavior trees method is based on the idea of incrementally building the model out of its building blocks, the functional requirements. This is accomplished by graphically representing each requirement as its own behavior tree and incrementally merging the trees to form a more complete behavior model of the system [23]. Fig. 18 (right) shows the fully integrated behavior tree model of the microwave oven.

R1 If the oven is idle and the user pushes the button, the oven will start cooking (that is, energize the power tube for one minute).
R2 If the button is pushed while the oven is cooking, it will cause the oven to cook for an extra minute.
R3 Pushing the button when the door is open has no effect (because it is disabled).
R4 Whenever the oven is cooking or the door is open, the light in the oven will be on.
R5 Opening the door stops the cooking.
R6 Closing the door turns off the light. This is the normal idle state prior to cooking when the user has placed food in the oven.
R7 If the oven times out, the light and the power tube are turned off and then a beeper emits a sound to indicate that the cooking is finished.

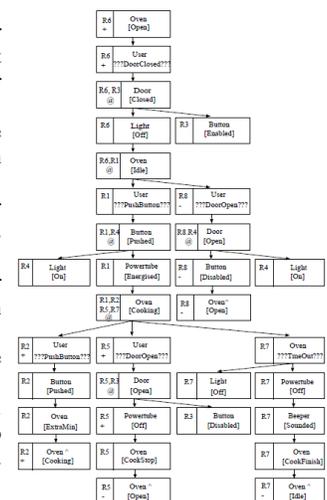

Fig. 18 The requirements of a microwave oven (left), and fully integrated behavior tree model of the microwave oven (right). From Winter et al. [23]

According to Dromey [9], conventional software engineering applies a design that will satisfy its set of functional requirements; in contrast to this, the behavior tree notation allows constructing a design out of its set of functional requirements, one at a time, into an evolving design behavior tree. This significantly reduces the complexity of the design process and any subsequent change process. The behavior tree designers can easily build complex behavior composing simpler ones, which represents a key advantage of behavior trees [24].

It seems that the behavior trees method utilizes a formal and "easy to understand tree-structured graphs, to represent functional requirements" [25]. Each requirement can be translated into one behavior tree and these behavior trees can be integrated into one single behavior tree called a design behavior tree. During a software system's evolution, a separate tree can represent each version of the system [25].

From the TM modeling point of view, such a method does not distinguish the fundamental distinction between potentiality and actuality.

In this paper, we introduce the TM modeling as an alternative approach to reach an integrated behavior with clear semantics, for example, definitions of actions, events and behavior. The TM modeling of behavior has its own advantages because behavior is an integral phase of the modeling process. This example provides both models of the microware system side by side for contrasting purposes.

To start the TM modeling, we need to introduce a special thimac called *full potentiality* to apply in this example. The potentiality thimac (not potentially at large) is a property of a system where everything (other thimacs) are not realized, that is, in the potential level. For example, a microware is initially in the state of full potentiality. No subpart of it is creating events (here, we ignore the mere existence). This type of thimac *full potentiality* (point (0,0) in the coordinate system of Fig. 8 (b)) is called a T-state by Lupasco, who described it as not in potentiality or in actuality, that is, neither false nor true. It represents a state outside potentiality or actuality. We use it as a *starting* condition of the microwave. Details of this issue will be delayed to further research.

Fig. 19 shows the static model of the microwave oven, described as follows, using the given requirements in Fig. 18.

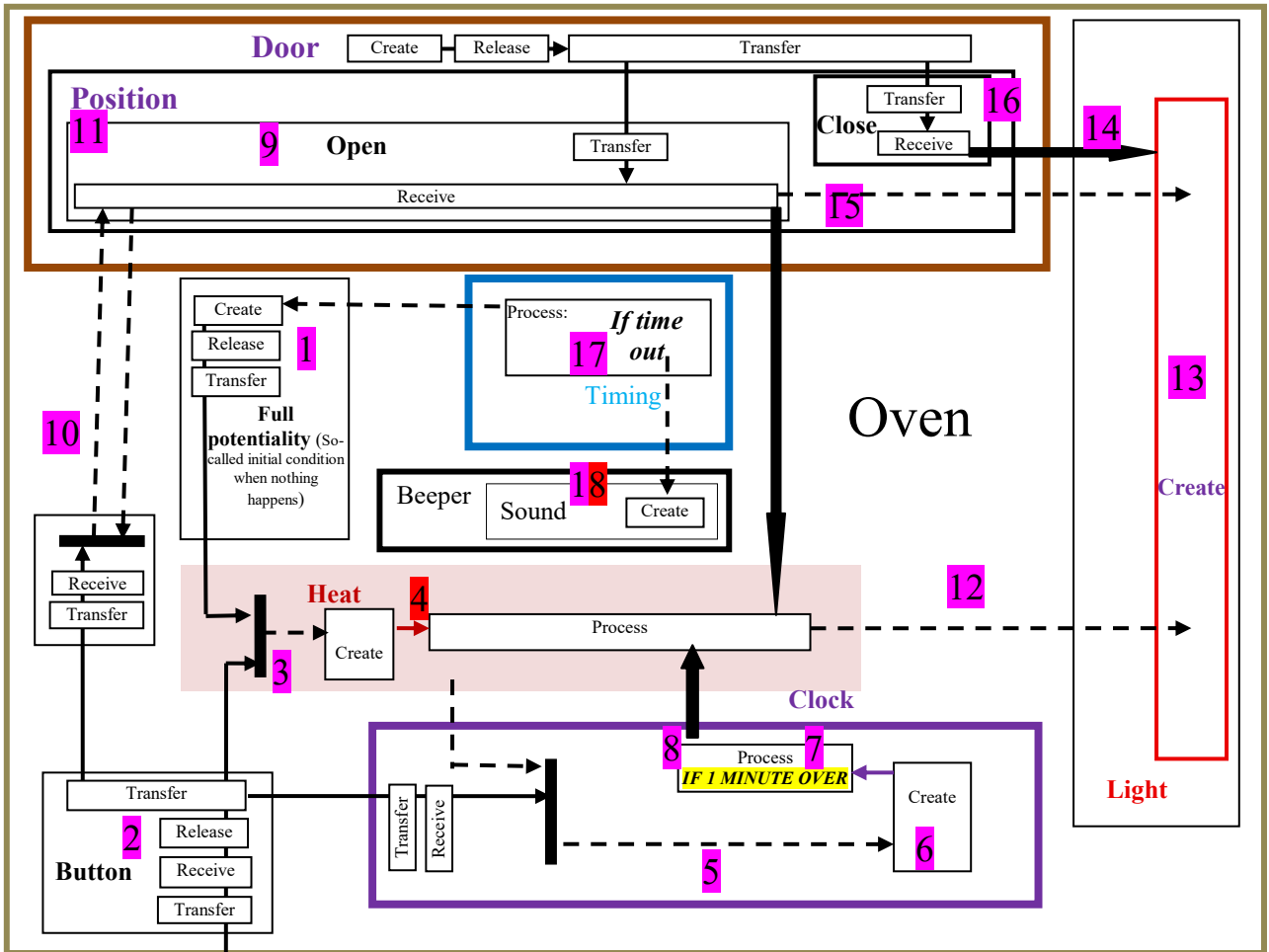

Fig. 19 The static TM model of the microwave oven

**R1**: The oven is idle (pink number 1). This is indicated by having full potentiality (no part of it is actual). Additionally, the button has been pushed (2). This would trigger (3) the oven to start heating/cooking (4). The vertical bar is a shorthand simplification that can be replaced with a thimac.

**R2**: The button is pushed (2) while the oven is creating heat (4) that triggers (5) the clock timing that is set to one minute (6). The time is processed (7) and if it is one minute, the heating is *stopped* (8). This is a negative event; hence, it will be implemented at the dynamic level (in Fig. 19 it is represented as a black arrow). Note that we change "oven is idle" in the original description because "idle" means returns to its initial condition, for example, closed door and such. According to our best guess, the authors meant turning off the heat.

**R3**: The button is pushed (2) while the door is open (9) triggers (10) *nothing* (nothing is in the original door position, 11). We model this nothing as keeping things as they are, thus opening a door that is already open.

**R4**: The oven is cooking/heating (4) triggers (12) creating light (13) in the bulb. The door is open (9) triggers creating light in the bulb (13).

**R5**: The door is open (9) triggers the oven heat off (15). This is a negative event; hence, it will be implemented at the dynamic level and indicated by a black arrow.

**R6**: The door is closed (16) triggers the negative event of returning the bulb to its static condition (14), which is represented by a black arrow.

**R7**: The oven timing out (17–assuming that it was set previously) triggers the beeper to create a sound (18) and the oven to be in its initial condition, that is, turned off (1).

Note that some of the given requirements are related to events at the dynamic level. Hence, these event-related requirements are marked at the static level by thick black arrows.

Fig. 20 shows the dynamic TM model, which involves the following events.

$E_1$: The oven is in full potentiality, that is, no events.
$E_2$: The user pushes the button.
$E_3$: The oven is heating.
$E_4$: The clock is set to one minute.
$E_5$: The clock time is processed.
$E_6$: The clock time reaches one minute.
$E_7$ (**NE$_3$**): Heating ($E_3$) is potentialized, that is, stopped.
$E_8$: The door is open.
$E_9$ (**NE$_3$**): Heating ($E_3$) is potentialized, that is, stopped.
$E_{10}$: The light is on.
$E_{11}$: The door is closed.
$E_{12}$ (**NE$_{10}$**): The light on ($E_{10}$) is potentialized, that is, off.
$E_{12}$: The oven times out.
$E_{13}$: The beeper makes a sound.

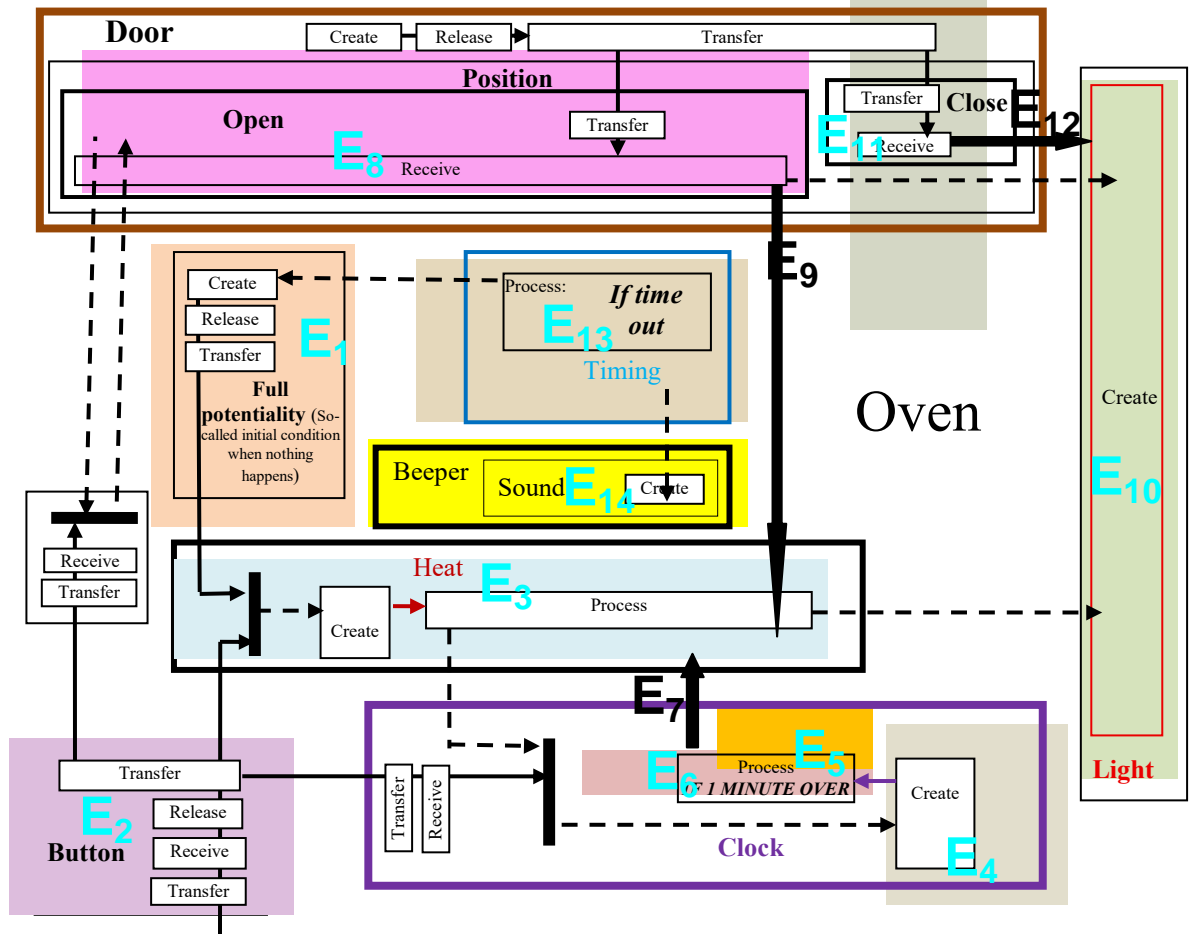

Fig. 20 The dynamic TM model of the microwave oven

Fig. 21 shows the integrated behavior of the microwave system.

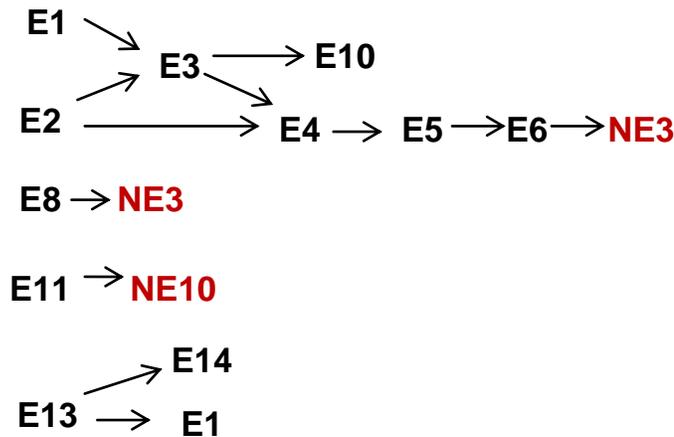

Fig. 21 Behavior model

## VI. CONCLUSION

TM modeling provides a framework for analyzing and explaining real-world entities and processes. This modeling does not pretend to achieve any the level of rigor theory as, for example, in logic and mathematics. The TM model is not intended to supersede any existing modeling approaches. These models are fundamentally based on similar underlying concepts, but they have a different focus, are represented using different modeling languages taken from different viewpoints, utilize different terminology and are used to develop different artefacts; therefore, they typically lack consistency and alignment [26]. TM modeling is proposed rather as a methodology that would expose conflicting modeling notions as produced and used in current models.

From the TM perspective, such study is essential to the overall development of understanding conceptual modeling.

TM components involve the following.
- Conceptual dualities of reality that include static and dynamic levels.
- Categorical structure of entities and actions.
- A diagrammatic representation to reflect the intended domain of modeling.

This paper extends the modeling sphere to reality through incorporating some underlying notions in the universe. The aim is to develop a reasonable level of ontological understanding that clarifies some of the ongoing issues on the nature and function of conceptual modeling. We incorporate the notion of non-contradiction in conceptual modeling using already available theoretical concepts such as Lupasco's non-negativity notion.

Specifically, we base the TM model on the two-level ontological levels construction of entities and actions. It seems that no current conceptual modeling in software engineering adopts such a view. It is too tedious to become involved in a philosophical discussion about this; instead, for now, we advocate putting the representation of each methodology side by side to evaluate the advantages/disadvantages of each presentation. It seems that TM modeling exhibits some benefits from the theoretical (semantics) point of view. Of course, philosophers have already digested these issues. As a negative note for the two-level hypothesis, according to Inwagen [27], there is no place for the concept of entities that occupy different ontological levels in a metaphysical theory that affirms the existence of only substances and abstract objects.